\documentclass[]{pasj01}
\usepackage{ulem}

\begin{document}

\title{Detection of highly correlated optical and X-ray variations in SS Cygni with Tomo-e Gozen and {\it NICER}}

\author{Yohei \textsc{Nishino}\altaffilmark{1}}%
\altaffiltext{1}{Department of Astronomy, Graduate School of Science, The University of Tokyo, 7-3-1 Hongo, Bunkyo-ku, Tokyo 113-0033, Japan}
\email{yohei.nishino@grad.nao.ac.jp}

\author{Mariko \textsc{Kimura}\altaffilmark{2}}
\altaffiltext{2}{Extreme Natural Phenomena RIKEN Hakubi Research Team, Cluster for Pioneering Research, RIKEN, 2-1 Hirosawa, Wako, Saitama 351-0198, Japan}

\author{Shigeyuki \textsc{Sako}\altaffilmark{3,4}}
\altaffiltext{3}{Institute of Astronomy, Graduate School of Science, The University of Tokyo, 2-21-1 Osawa, Mitaka, Tokyo 181-0015, Japan}
\altaffiltext{4}{The Collaborative Research Organization for Space Science and Technology, The University of Tokyo, 7-3-1Hongo, Bunkyo-ku, Tokyo 113-0033, Japan}

\author{Jin \textsc{Beniyama}\altaffilmark{3}}
\author{Teruaki \textsc{Enoto}\altaffilmark{2}}

\author{Takeo \textsc{Minezaki}\altaffilmark{3,4}}
\author{Nozomi \textsc{Nakaniwa}\altaffilmark{5}}
\altaffiltext{5}{Department of Physics, Tokyo Metropolitan University, 1-1 Minami-Osawa, Hachioji, Tokyo 192-0397, Japan}
\author{Ryou \textsc{Ohsawa}\altaffilmark{3}}
\author{Satoshi \textsc{Takita}\altaffilmark{3}}
\author{Shinya \textsc{Yamada}\altaffilmark{6}}
\altaffiltext{6}{Department of Physics, Rikkyo University, Toshima-Ku, Tokyo, 171-8501, Japan}
\author{Keith C. \textsc{Gendreau}\altaffilmark{7}}
\altaffiltext{7}{NASA Goddard Space Flight Center, Greenbelt, MD 20771, USA}

\KeyWords{accretion, accretion disks --- novae, cataclysmic variables --- stars: dwarf novae --- stars: individual (SS Cyg)}

\maketitle

\begin{abstract}
We report on simultaneous optical and X-ray observations of the dwarf nova SS Cyg with Tomo-e Gozen/1.05 m Kiso Schmidt and Neutron star Interior Composition ExploreR ({\it NICER}) / International Space Station (ISS). A total of four observations were carried out in the quiescent state and highly correlated light variations between the two wavelengths were detected. We have extracted local brightness peaks in the light curves with a binning interval of 1 sec, called `shots', and have evaluated time lags between the optical and X-ray variations by using a cross-correlation function.
Some shots exhibit significant optical lags to X-ray variations and most of them are positive ranging from $+$0.26 to 3.11 sec, which have never been detected.
   They may be ascribable to X-ray reprocessing in the accretion disk and/or the secondary star. This analysis is possible thanks to the high timing accuracy and the high throughput of {\it NICER} and the matching capability of Tomo-e Gozen. Also, it is confirmed that the correlation between the optical and X-ray variations changed in the middle of one of our observation windows and the X-ray spectrum softer than 2 keV varied in accordance with the correlation.
\end{abstract}
\section{Introduction}

There is a wide variety of accreting objects in the universe, including cataclysmic variables (CVs), X-ray binaries (XBs), young stellar objects, and active galactic nuclei (AGNs).  They exhibit similar transient phenomena and fast variability due to the accretion from the disk to the central object (\cite{1995cvs..book.....W}; \cite{2008bhad.book.....K}; \cite{2007A&ARv..15....1D}).
Fast stochastic light variability is often seen in the accreting objects over wide wavelengths (e.g., \cite{1999MNRAS.309..803G}; \cite{2004astro.ph.10551V}).
It is considered that stochastic variability originates from fluctuations in mass accretion rates, which are generated at different disk radii and propagate inwards (\cite{1997MNRAS.292..679L}; \cite{1997ApJ...486..388Y}).
Multi-wavelength studies have been performed to reveal mechanisms of the stochastic variability and structures of the inner accretion flows (e.g., Balman \& Revnivtsev \yearcite{2012A&A...546A.112B}; \cite{2017MNRAS.468.1183D}).

Dwarf novae (DNe), one subclass of CVs, are close binary systems composed of a white dwarf (WD) surrounded by an accretion disk and a low-mass star called the secondary star (see \cite{1995cvs..book.....W} for a review).  They experience intermittent outbursts with a duration of weeks-months due to thermal-viscous instability exerted in the disk (\cite{1996PASP..108...39O}) and show stochastic variability with large amplitudes in the quiescent state (Bruch \yearcite{1992A&A...266..237B}, \yearcite{2021MNRAS.503..953B}).
DNe are one of the best laboratories for studying the mechanisms of stochastic variability because we have a large number of bright targets and can easily obtain their light curves on timescales of seconds to minutes.  Most of the optical radiation comes from the accretion disk, the secondary star, and WD. 
It is considered that X-ray emitting hot plasma exists in the vicinity of the WD (Narayan \& Popham \yearcite{1993Natur.362..820N}; Patterson \& Raymond \yearcite{1985ApJ...292..535P}; \cite{2015AcPPP...2..116B}, \yearcite{2020AdSpR..66.1097B}).

In this letter, we study the brightest dwarf nova SS Cyg.  This system normally shows outbursts with intervals of about 1 month and with amplitudes of about 3.5 mag at optical wavelengths and the quiescent level is around 12 mag in the optical $V$ band.  However, the quiescent brightness has begun increasing since August 2019 at optical and X-ray wavelengths \citep{2020ATel13744....1N}.  We considered that this is the best opportunity for observing the fast stochastic variability in this system and were engaged in coordinated observations with Tomo-e Gozen \citep{2018SPIE10702E..0JS} and Neutron star Interior Composition ExploreR ({\it NICER}) \citep{2016SPIE.9905E..1HG}.

This letter focuses on highly correlated fast variations between optical and X-ray wavelengths in SS Cyg and may provide some implications for interpreting the ongoing anomalous events in this system.  This letter is structured as follows.  In section 2, our methods of simultaneous observations with the optical and X-ray instruments are described. In section 3, we present the obtained light curves and our analysis results. In section 4, the origins of the time lags and high correlations that we detected are discussed.

\section{Observations}\label{Observations}
Tomo-e Gozen is a wide-field video camera with 84 chips of CMOS image sensors for the 1.05 m Kiso Schmidt telescope \citep{2018SPIE10702E..0JS}. This camera is capable of obtaining consecutive frames with timestamps of 0.2 msec absolute accuracy. The frame rate can be increased by reducing the fields-of-views of the sensors. The simultaneous observations were carried out with a single chip of the CMOS sensors at 26 fps achieving a field-of-view of 108 arcsec × 44 arcsec. Monochromatic imaging was carried out without a filter. The sensitive wavelengths are between 380 and 710 nm.

   {\it NICER} is an X-ray instrument onboard the International Space Station, which is composed of 56 pairs (52 sets are now active on orbit) of silicon drift detectors and concentrator optics with high efficiency between 0.2 and 12 keV. The large effective area in soft X-rays (1900~cm$^2$ at 1.5 keV) of the {\it NICER} optics allows us to study fast variability of X-ray objects thanks to its high throughput. The absolute time tagging accuracy is better than 300 nsec. The high absolute timing accuracy of both instruments enables us to evaluate lags between optical and X-ray light variations (see section 3.2).

\begin{table*}[h]
  \caption{Observation logs of our simultaneous observations. Observation IDs are allocated to the {\it NICER} data.}
  \label{log_table}
  \centering
  \scalebox{0.9}{
  \begin{tabular}{lccccc}
    \hline\noalign{\vskip3pt}
\multicolumn{1}{c}{ObsIDs} & Date & Start (UTC) & Stop (UTC) & Duration (sec) & Window \\ [2pt]
\hline\noalign{\vskip3pt}
 3201600128 & September 14th, 2020  & 10:11:30 & 10:20:29 & 539 & W1\\
 3201600129 & September 15th, 2020 & 12:28:26 & 12:40:41 & 735 & W2\\
 3201600146 & November 14th, 2020 & 12:06:28 & 12:47:20 & 2452 & W3\\
 3201600148 & November 18th, 2020 & 10:59:57 & 11:16:57 & 1020 & W4\\
    \hline
  \end{tabular}
  }
\end{table*}

   Optical light curves of SS Cyg and a neighbor reference star were derived from video frames of Tomo-e Gozen with a standard photometry method. The aperture radius was fixed to 18 arcsec, and the apertures were set at the centroids of targets. Time variations of measured flux densities due to atmospheric fluctuation were corrected with the relative photometry method with a bright source in the same frame. The photometric zero points of the Tomo-e Gozen observations with no filter were derived using the G-band magnitude of the {\it Gaia} EDR3 catalog (\cite{2016A&A...595A...2G}, \yearcite{2021A&A...649A...1G}), since the photometric system of the {\it Gaia} G-band has a similar wavelength dependence to that of Tomo-e Gozen.

   We utilized HEAsoft version 6.27.2 for data reduction and analysis of the {\it NICER} data. The data were reprocessed with the pipeline tool \texttt{nicerl2} based on the {\it NICER} CALDB version later than 2019 May 16, before producing light curves of the individual ObsIDs. We extracted the 0.3--7~keV light curves by using \texttt{lcurve}.  
   The observation times of both optical and X-ray light curves are converted to barycentric Julian date (BJD) in terrestrial time (TT), which is a time standard for observations from the Earth defined by International Astronomical Union. The observation logs are shown in table \ref{log_table} and we denote the observation windows on September 14th and 15th, November 14th and 18th in 2020 as W1, W2, W3, and W4, respectively.

\section{Results}\label{Results}

\subsection{Light curves}
Light curves of SS Cyg at optical and X-ray wavelengths taken in the optical quiescence state are shown in figure 1. The optical light curves are resampled to 5 sec and the X-ray events are grouped for each optical time bin.

\begin{figure}[h]
  \begin{center}
  \includegraphics[width=8cm]{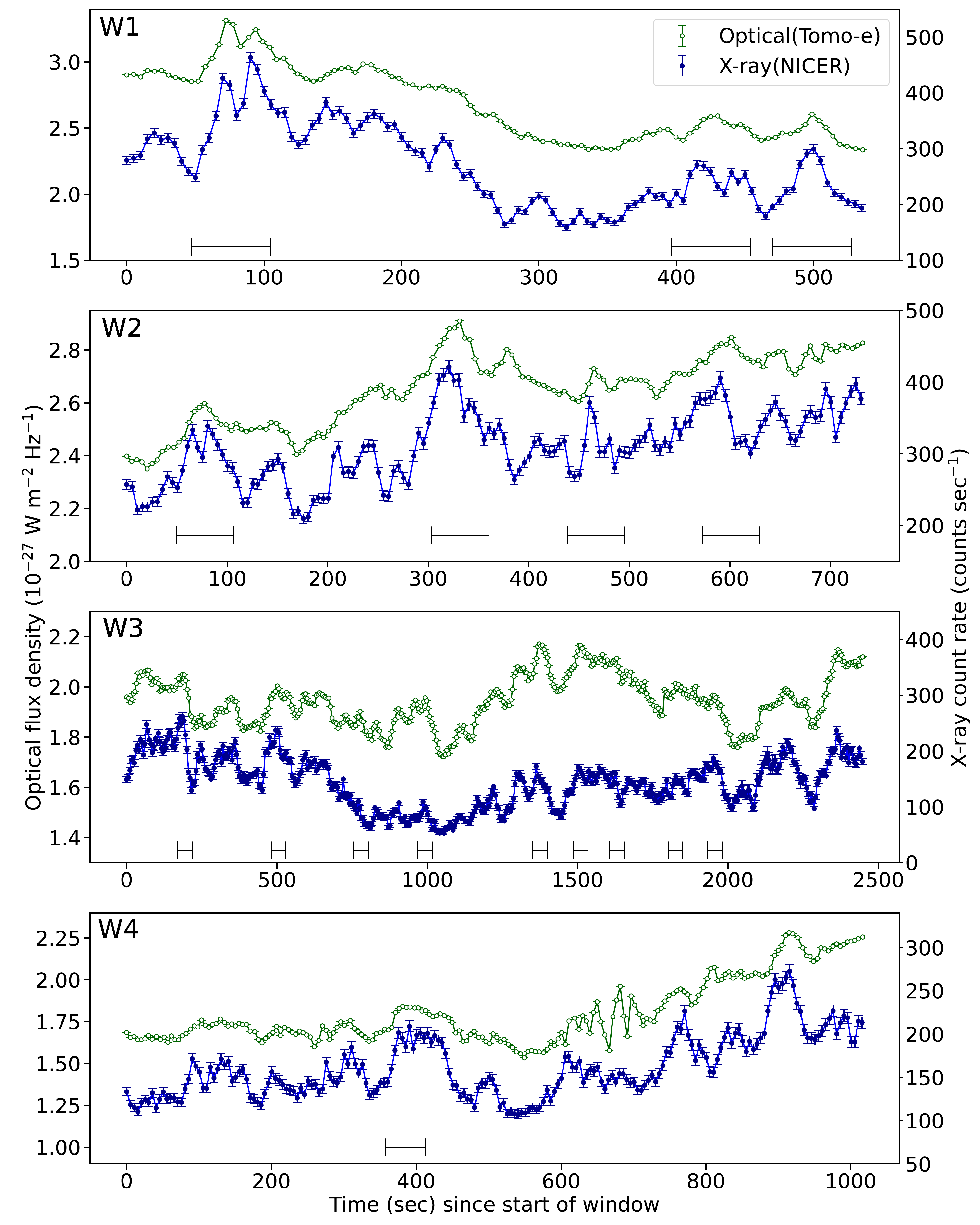}
  \end{center}
  \caption{Light curves of SS Cyg at optical (green) and X-ray (blue) wavelengths in the observation windows on September 14th (W1), September 15th (W2), November 14th (W3) and November 18th (W4) in 2020, respectively. Both of these light curves have 5 sec bin time.} Each sample has error bars corresponding to photometric errors. Time windows used for evaluations of time lags between optical and X-ray variations in subsection \ref{Shot lag analysis} are indicated by horizontal bars.
  \label{all_light_curve}
\end{figure}

   Cataclysmic brightness changes with large amplitudes and short timescale variations are found in the light curves (see figure \ref{all_light_curve}). The most interesting thing is that amplitude variations and their phases of the light curves at optical and X-ray wavelengths are quite similar to each other in all observation windows. Such clear correlations of brightness changes between optical and X-ray wavelengths have never been reported in SS Cyg previously.

\subsection{Shot lag analysis}\label{Shot lag analysis}

Time lags between the optical and X-ray light curves were evaluated by using the shot-profile method (\cite{1994ApJ...423L.127N}; \cite{2019A&A...631A.134D}) and the cross-correlation function (CCF) method (e.g., \cite{1998PASP..110..660P}).The optical light curves are resampled to 1 sec and the X-ray events are grouped for each optical time bin.

   We first searched for local brightness peaks in the optical light curve with a time width of 60 sec and defined a set of time series data around the local peak with a time width of 30 sec as a shot to avoid overlapping between adjacent shots. The length of one shot is set to 60 sec to evaluate lags on 1 sec time scale between shots on 1 minute time scale. The typical duration of a single fluctuation is 60 sec, and we confirmed that a slight change of the time width by $\pm$ 20 sec does not significantly change the results.

\begin{figure}[htb]
  \centering
  \includegraphics[width=6cm]{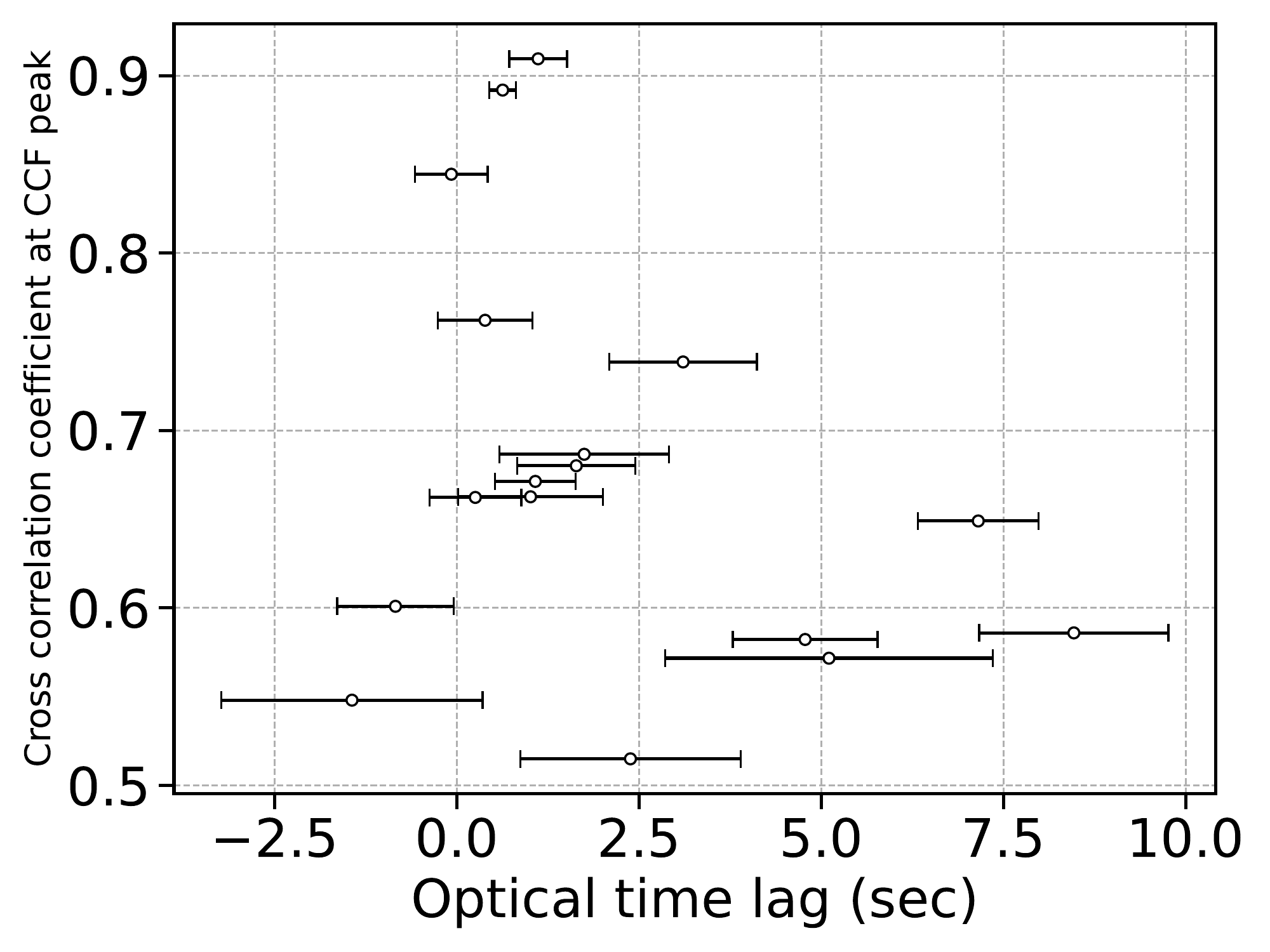}
  \caption{Optical time lags against X-ray and cross-correlation coefficients at CCF peaks. Seventeen shots are extracted from the data sets in the four observation windows. The time lags are distributed within a range of $+$0.26 to $+$3.11 sec when the cross-correlation coefficients at CCF peaks are larger than 0.65. The errors are calculated from 1000 FR/RSS realizations (see text).
  }
  \label{Lag_max_Correlation_1err}
\end{figure}

   We next calculated a CCF between the optical and X-ray variations
   for each shot by the interpolated CCF method as applied in
   White \& Peterson (\yearcite{1994PASP..106..879W}).
   Taking into account the dependence of the binned X-ray data on adjacent points in the resampling process, we set the criterion of the maximum of the CCFs to 0.50, though CCF values with the sample size $N=60$ higher than 0.36 are statistically significant at a confidence level of 99 \%. We selected a total of 17 shots as shown by horizontal bars in figure \ref{Lag_max_Correlation_1err}. We show all the values of the lags with error bars in table \ref{lag_table} and their CCFs are displayed in figure E1. We confirmed that the ACFs are featureless (see figure E2 and E3).
   
   We defined the time lag between the optical and X-ray light curves, $\tau_{\mathrm{cent}}$, by the centroid around the CCF peak that is computed from all neighboring data points with cross-correlation coefficients larger than 0.8 times the peak value of the CCF, as conventionally adopted in the reverberation mapping studies of AGNs (e.g., \cite{1998PASP..110..660P}). The 1$\sigma$ uncertainties are calculated by the Monte Carlo simulations of flux randomization (FR) and random subset sampling (RSS) (Peterson et al. \yearcite{1998PASP..110..660P}, \yearcite{2004ApJ...613..682P}).
   
   Some shots exhibit significant lags. All shots except one with cross-correlation coefficients at CCF peaks larger than 0.65 have optical time lags distributed within a range of $+$0.26 to $+$3.11 sec.

\begin{table}[h]
  \caption{Properties of all the shots.}
  \label{lag_table}
  \centering
  \scalebox{0.8}{
  \begin{tabular}{lcccc}
    \hline\noalign{\vskip3pt}
\multicolumn{1}{c}{Window} & Start\footnotemark[$\dagger$] (sec) & End\footnotemark[$\ddagger$] (sec) & CCF max & lag (sec)  \\ [2pt]
\hline\noalign{\vskip3pt}
W1 & 46.0	& 106.0 &	0.89 &	$0.62 \pm 0.19$ \\
   & 395.0	& 455.0 &	0.65 &	$7.11 \pm 0.82$ \\
   & 469.0  &	529.0 &	0.85 &	$-0.07 \pm	0.47$ \\
W2 & 48.0 &	108.0 &	0.66 & 	$0.26 \pm 0.64$ \\
   & 302.0 &	362.0 &	0.69 &	$1.67 \pm 1.08$ \\
   & 437.0 &	497.0 &	0.67 &	$1.10 \pm 0.55$ \\
   & 571.0 &	631.0 &	0.59 &	$8.46 \pm 1.34$ \\
W3 & 163.0 &	223.0 &	0.91 &	$1.11\pm0.39$ \\
   & 475.0 &	535.0 &	0.51 &	$2.36\pm1.48$ \\
   & 749.0 &	809.0 &	0.57 &	$5.04\pm2.20$ \\
   & 962.0 &	1022.0 &	0.68 & $1.63\pm	0.82$ \\
   & 1344.0 &	1404.0 &	0.57 & $4.75\pm	0.98$ \\
   & 1480.0 &	1540.0 &	0.67 &	$0.95\pm0.91$ \\
   & 1600.0 &	1660.0 &	0.73 &	$3.11\pm1.01$ \\
   & 1795.0 &	1855.0 &	0.54 &	$-1.49\pm1.77$ \\
   & 1926.0 &	1986.0 &	0.60 &	$-0.85\pm0.66$ \\ 
W4 & 355.0 &	415.0 &	0.76 & $0.39\pm0.65$ \\
    \hline
  \end{tabular}
  }
\begin{tabnote}
{\hbox to 0pt{\parbox{170mm}{\footnotesize
\par\noindent
\hangindent5pt\noindent
\hbox to6pt{\,\footnotemark[$\dagger$]\hss}\unskip%
Start time of each shot. \\
\noindent {\footnotemark[$\ddagger$]\hss}\unskip%
End time of each shot.\\
\noindent {\footnotemark[$\ddagger$]\hss}\unskip%
Time delay of optical variations against X-ray ones
}\hss}}
\end{tabnote}
\end{table}

\subsection{Correlation of variations}\label{Linear correlations}
\begin{figure}[h]
  \centering
  \includegraphics[width=8cm]{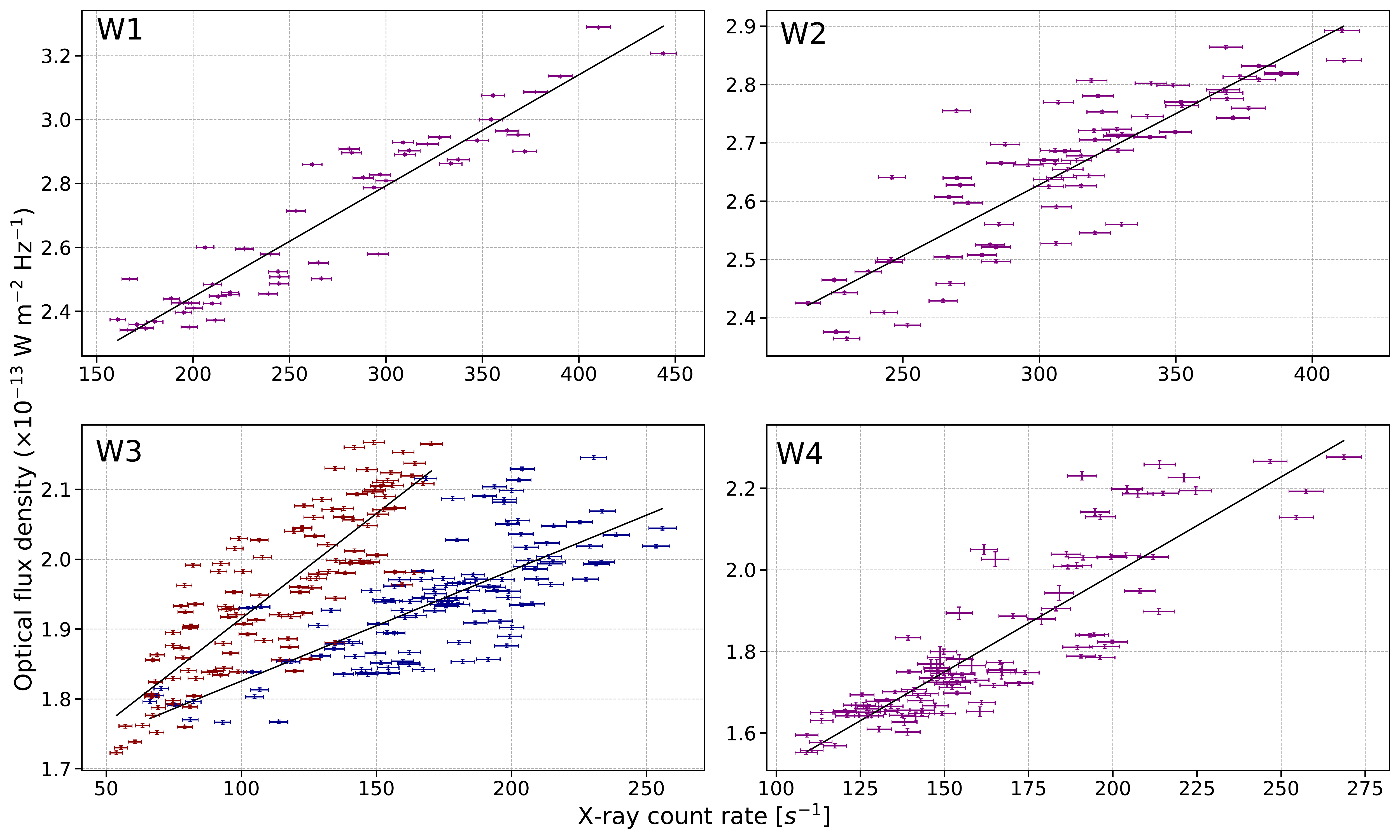}
  \caption{Scatter plots of optical and X-ray fluxes in the W1, W2, W3, and W4 observation windows. Fitted linear functions are shown with bold lines over the plots.}
  \label{Linear}
\end{figure}

Scatter plots of the optical flux density and X-ray count rate in the four observation windows are shown in figure \ref{Linear}. Positive correlations are clearly seen except for the plot in W3. We evaluated the correlations with a simple linear function;
\begin{equation}\label{eq:one}
    F_{\mathrm{opt}} = AF_{\textrm{X-ray}} + B, 
\end{equation}
using the optical light curves resampled to 10 sec and the X-ray light curves grouped for each optical time bin.
The coefficients of determination, $R^2$, are listed in table \ref{tab:second}.
 The linear functions with the fitted parameters are drawn with bold lines in figure \ref{Linear}. The flux variations in the optical and X-ray wavelengths are highly correlated in W1, W2, and W4. There seem to be two types of correlations between the optical and X-ray fluxes in W3.
We fit them independently and show the correlation coefficients in table \ref{tab:second}.

\begin{table}[h]
  \caption{The coefficients of determination and power-law indexes}
  \label{tab:second}
  \centering
  \scalebox{0.9}{
  \begin{tabular}{lccc}
    \hline\noalign{\vskip3pt}
\multicolumn{1}{c}{Window} & $R^2$\footnotemark[$\dagger$] & $\alpha_{\rm{Opt}}$\footnotemark[$\ddagger$] & $\alpha_{\rm{X-ray}}$\footnotemark[$*$] \\ [2pt]
\hline\noalign{\vskip3pt}
 W1 & 0.89 & -1.59(0.26) & -1.52(0.37) \\
 W2 & 0.75 & -1.90(0.27) & -2.26(0.32) \\
 W3 & 0.56\footnotemark[$\S$] & -1.98(0.27) & -1.62(0.16) \\
    & 0.71\footnotemark[$\P$] &  & \\
 W4 & 0.77 & -1.75(0.14) & -1.88(0.15) \\
    \hline
  \end{tabular}
  }
\begin{tabnote}
{\hbox to 0pt{\parbox{170mm}{\footnotesize
\par\noindent
\hangindent5pt\noindent
\hbox to6pt{\,\footnotemark[$\dagger$]\hss}\unskip%
The coefficients of determination for linear correlation.  \\
\noindent {\footnotemark[$\ddagger$]\hss}\unskip%
Power-law index for the optical power spectrum. \\
\noindent {\footnotemark[$*$]\hss}\unskip%
Power-law index for the X-ray power spectrum. \\
\hbox to6pt{\,\footnotemark[$\S$]\hss}\unskip%
$t<700, 1900<t\ \rm{sec}$. \\
\noindent {\footnotemark[$\P$]\hss}\unskip%
$700<t<1900\ \rm{sec}$.
}\hss}}
\end{tabnote}
\end{table}

\subsection{Power spectrum density}\label{Power spectrum}
\begin{figure}[h]
  \centering
  \includegraphics[width=7cm]{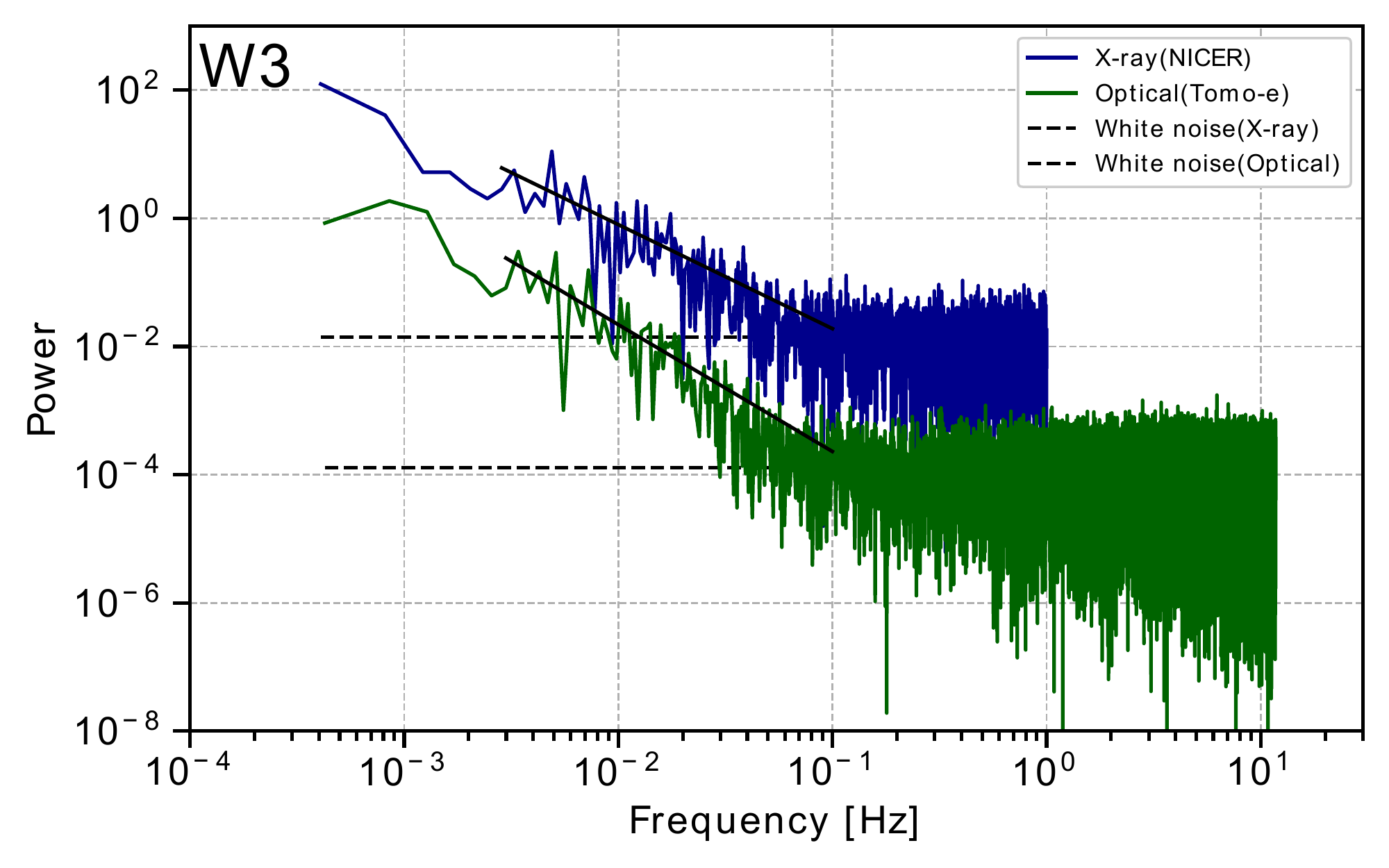}
  \caption{Power spectral density of X-ray (blue) and optical (green) light curves in W3. The fitted results and white noise are shown with bold and dashed lines, respectively. Several points in low frequencies are omitted due to low statistics.}
  \label{PSD}
\end{figure}

We obtained power spectral densities (PSDs) of the optical and X-ray light curves. 
The light curves are normalized by mean fluxes. 
A PSD in W3 is displayed in figure \ref{PSD} and those in the other windows are displayed in figure E2. 
Neither periodic variability nor QPO were detected in the optical quiescence in SS Cyg in both wavelengths.
We fitted PSDs with power law and evaluated slopes of the power at higher frequencies than 0.1 Hz. The lowest frequencies depend on the lengths of each data, but these are approximately after the frequency breaks described in previous studies (e.g., Balman \& Revnivtsev \yearcite{2012A&A...546A.112B}).
The best-fitted power-law indexes are listed in table \ref{tab:second}. 
We also show the Poisson noise level estimated by a mean power at higher frequencies than 0.1 Hz. 
The power-law indexes in optical and X-ray wavelengths imply that the frequency range is dominated by red noise, $1/f^2$, consistent with previous studies (e.g., \cite{2018MNRAS.481.2140A}; \cite{2015AcPPP...2..116B}, \yearcite{2020AdSpR..66.1097B}).

\section{Discussion}\label{Discussion}
Balman \& Revnivtsev (\yearcite{2012A&A...546A.112B}) reported that a cross-correlation coefficient between the X-ray and ultraviolet (UV) light curves of $\sim$0.4 at most and an X-ray delay of 166--181 sec were measured in simultaneous X-ray and UV observations of SS Cyg in the optical quiescence. They interpreted it as the propagation time of matter from a truncated inner disk to the WD surface.
On the other hand, we focus on shorter timescales. Our analysis of the shot lags indicates that the cross-correlation coefficients between the short-term optical and X-ray variations peak at 0.9. (see subsection \ref{Shot lag analysis}). 

As we have shown in section 3.2, some shots exhibit significant lags ranging from $+$0.26 to 3.11 sec. Such positive lags have never been reported in SS Cyg.
Furthermore, as shown in the left panels of figure \ref{Linear}, the good linear correlations between the optical and X-ray variations are found in the entire light curves in W1, W2, and W4, though the correlation in W3 is exceptional as discussed later. These results indicate the X-ray reprocessing model, in which the X-ray emission from the vicinity of the WD irradiates the disk and/or the secondary star, and is reprocessed into the optical emission.

However, the timescale of the optical lags is still controversial \citep{2013MNRAS.431.2535S}. The optical lags of +0.26--3.11 sec with high CCF peaks are consistent with the light travel time from the boundary layer to the outermost radius of the disk of SS Cyg, about $5\times10^{10}~\mathrm{cm}$ \citep{1992PhDT.......119S, 2017MNRAS.472.2937H}, or to the surface of the secondary star. On the other hand, \citet{2018MNRAS.481.2140A} reported the soft lags of $\sim$ 5 sec in the optical bands that the redder emission lags the bluer emission, and explained them by the thermal timescale that is required to alter the local thermal equilibrium on the surface of the disk by the irradiation. Since the X-ray-to-optical lags are considered to be similar to or slightly larger than the soft lags in the optical bands under the X-ray reprocessing model, the measured optical lags, which are larger than the timescale of the light travel time, are also consistent with the thermal timescale.
   
   The delay of optical variations relative to X-ray variations has never been reported in SS Cyg previously, though there is a possibility that previous observations did not have the sensitivity to detect short time lags.  Our results may suggest that X-ray reprocessing of the disk and/or the secondary star recently became stronger in this system during our observations. \citet{2021PASJ..tmp...85K} suggested that the X-ray emitting region in SS Cyg vertically expanded during the optical quiescence in 2020 through their broadband X-ray spectral analyses.  An increase in the scale height of the X-ray emitting region may link to stronger X-ray reprocessing and our results may support the interpretation in \citet{2021PASJ..tmp...85K}.
   
   The optical and X-ray relationship in W3 is somehow different from those in the other observation windows. The linear correlation is roughly divided into two types in W3 (see figure \ref{Linear}). The red and blue points in figure \ref{Linear} correspond to the points with the same colors in the light curves in the left panel of figure E3.  The correlation seems to change in the middle of this time window (from 700 sec to 1900 sec). We analyze the {\it NICER} spectrum in W3 and find that the spectrum less than 2~keV also changed (see the right panel of figure E3). The absorption increased and the reflection effect might become stronger from 700 sec to 1900 sec (see also table E1). This result implies that a larger part of the X-ray emitting region was shielded by absorbers from 700 to 1900 sec, which could affect X-ray reprocessing.  Perhaps, the X-ray emitting region became more compact at that time.  Also, the shot-lag analysis revealed a significant optical lag before 700 sec.  It may be suggested that X-ray reprocessing occurred before soft X-rays were attenuated. Long-term, $\sim$10000 sec, and broadband observations are needed to investigate the nature of this event.
   In figure E4, we show the SED in W1 for comparison. We see that the same spectral model is applicable for the brighter state.


\begin{ack}
This research has been partly supported by Japan Society for the Promotion of Science Grants-in-Aid for Scientific Research (KAKENHI) Grant Numbers 21H04491, 16H06341, 18H01272, 18H01261, 18H04575, 18H05223, 20H01942, 20K11935, 20K22374 (MK), 20H01941 (SY) and 17H06363. This research is also supported in part by the Research Center for the Early Universe (RESCEU) of the School of Science at the University of Tokyo and the Optical and Near-infrared Astronomy Inter-University Cooperation Program. M. Kimura acknowledges support by the Special Postdoctoral Researchers Program at RIKEN.
   This work has also made use of data from the European Space Agency (ESA) mission
{\it Gaia} (https://www.cosmos.esa.int/gaia), processed by the {\it Gaia}
Data Processing and Analysis Consortium (DPAC
https://www.cosmos.esa.int/web/gaia/dpac/consortium). Funding for the DPAC
has been provided by national institutions, in particular the institutions
participating in the {\it Gaia} Multilateral Agreement.

\end{ack}

\section*{Supplementary data}

The following supplementary data is available at PASJ online.
Table E1 and figures E1--E4.



\end{document}